\documentclass[aps,prl,amsmath,amssymb,preprint]{revtex4-1}
\pdfoutput=1
\usepackage{graphicx}
\usepackage[version=3]{mhchem} 
\usepackage{dcolumn}
\usepackage{bm}

\begin{document}

\title{Self-regulation mechanism for charged point defects in hybrid halide perovskites}


\author{Aron Walsh*}
\email[Electronic mail:]{a.walsh@bath.ac.uk}
\affiliation{Centre for Sustainable Chemical Technologies and Department of Chemistry, University of Bath, Claverton Down, Bath BA2 7AY, UK}

\author{David O. Scanlon*}
\email[Electronic mail:]{d.scanlon@ucl.ac.uk}
\affiliation{University College London, Kathleen Lonsdale Materials Chemistry, Department of Chemistry, 20 Gordon Street, London WC1H 0AJ, UK}
\affiliation{Diamond Light Source Ltd., Diamond House, Harwell Science and Innovation Campus, Didcot, Oxfordshire OX11 0DE, UK}

\author{Shiyou Chen}
\affiliation{Key Laboratory of Polar Materials and Devices (MOE), East China Normal University, Shanghai 200241, China}

\author{X. G. Gong}
\affiliation{Key Laboratory for Computational Physical Sciences (MOE) and Surface Physics Laboratory, Fudan University, Shanghai 200433, China}

\author{Su-Huai Wei}
\affiliation{National Renewable Energy Laboratory, Golden, CO 80401, USA}

\date{\today}

\begin{abstract}
Hybrid halide perovskites such as methylammonium lead iodide (\ce{CH3NH3PbI3}) exhibit unusually low free carrier concentrations despite being processed at low-temperatures from solution. We demonstrate, through quantum mechanical calculations, that the origin of this phenomenon is a prevalence of ionic over electronic disorder in stoichiometric materials. Schottky defect formation provides a mechanism to self-regulate the concentration of charge carriers through ionic compensation of charged point defects. The equilibrium charged vacancy concentration is predicted to exceed 0.4\% at room temperature. This behaviour, which goes against established defect conventions for inorganic semiconductors, has implications for photovoltaic performance. 
\end{abstract}


\maketitle

Hybrid halide perovskites recently transitioned from chemical curiosities to solar energy champions.\cite{mcgehee2013materials}
They can be produced at low-temperature and low-cost to produce high light-to-electricity conversion efficiencies.\cite{snaith-1,snaith-6156,gratzel-6156,bhachu2015scalable,C4CC05231E} 
The highest efficiency devices are based on \ce{CH3NH3PbI3} (denoted here as MAPI), which consists of a singly-charged closed-shell methylammonium cation (\ce{CH3NH3+} or MA) at the centre of a \ce{PbI3-} cage formed of corner sharing octahedra. 
The same structure is adopted by the chloride and bromide perovskites, with solid-solutions on the anion sub-lattice also reported.\cite{weber1978ch3nh3pbx3,hao2014lead}

The defect chemistry and physics of inorganic perovskites have been well studied for almost a century, but they remain a complex case in solid-state science, with contributions from electronic disorder (delocalised and localised charges) and ionic point defects, as well as extended dislocations and grain boundaries.\cite{maier2004physical,smith-2000}
In contrast, little is known about the hybrid perovskites. 
Preliminary reports have demonstrated the shallow nature of common point defects, which can contribute to effective electron and hole generation or recombination.\cite{yin-063903,agiorgousis2014strong,buin2014materials}
An anomaly is that despite exceptionally low defect formation energies, the measured carrier concentrations of thin-films are also remarkably low, in the region of 10$^{9}$--10$^{14}$cm$^{-3}$,\cite{stoumpos2013semiconducting,laurie} and bulk electron-hole recombination is highly suppressed. 
In comparison, for pristine semiconductors (e.g Si and Ge) values of 10$^{10}$--10$^{13}$cm$^{-3}$ are common due to a combination of high-purity materials and large point defect formation energies, 
while in solution processed multi-component materials (e.g. \ce{Cu2ZnSnS4}) values of 10$^{16}$--10$^{18}$cm$^{-3}$ are frequently observed due to lower purity samples and smaller defect formation energies.\cite{luque2011handbook}

For inorganic perovskites, Schottky disorder is a dominant type of defect, which is associated with the formation of stoichiometric amount of anion and cation vacancies, which can be distributed randomly in a crystal.\cite{kroger-1974}
It is found, for example, in \ce{SrTiO3} and \ce{BaTiO3}. 
Following the notation of Kr\"oger and Vink, for methylammonium lead iodide, we can consider both `full' Schottky disorder
\begin{equation}\label{d1}
\textit{nil} \rightarrow \ce{V_{MA}^/} + \ce{V_{Pb}^{//} + 3\ce{V_I}^{\bullet}} + \ce{MAPbI3}
\end{equation}
and `partial' disorder with respect to the methylammonium iodide
\begin{equation}\label{d2}
\textit{nil} \rightarrow \ce{V_{MA}^/} + \ce{V_I}^{\bullet} + \ce{MAI}
\end{equation}
or lead iodide
\begin{equation}\label{d3}
\textit{nil} \rightarrow \ce{V_{Pb}^{//} + 2\ce{V_I}^{\bullet}} + \ce{PbI2}
\end{equation}
sub-lattices.
Reactions \ref{d1}, \ref{d2} and \ref{d3} are charge-neutral (self-compensated), e.g. 
$[\ce{V_{MA}^/}]+[\ce{V_{Pb}^{//}}] \equiv 3 [\ce{V_I}^{\bullet}]$
for \ref{d1}.
While the individual isolated point defects have a net charge, the sum of these charges is zero for a macroscopic sample,
and does not involve the generation of electron or hole carriers.
Reaction \ref{d1} preserves the overall stoichiometry of the material, but Reactions \ref{d2} and \ref{d3} result in the loss of \ce{MAI} ($\mu_{\ce{CH3NH3}} + \mu_{\ce{I}} = \Delta H_{f(\ce{CH3NH3I})}$) and \ce{PbI2} ($\mu_{\ce{Pb}} + 2 \mu_{\ce{I}} = \Delta H_{f(\ce{PbI2})}$), respectively, and can be associated with non-stoichiometry.

The equilibrium concentration of lattice vacancies arising from Schottky disorder can be calculated by applying the law of mass action to Reaction \ref{d1}:
\begin{equation}\label{d4}
[\ce{V_{MA}^/}][\ce{V_{Pb}^{//}][\ce{V_I}^{\bullet}}]^3 = K_C = K_C^{\circ} exp \left(\frac{- \Delta H_S}{k_b T} \right),
\end{equation}
where $K_C$ represents the fraction of the lattice sites ($K_C^{\circ}$) that are vacant due to the reaction enthalpy ($\Delta H_S$), and contributions from the changes in vibrational entropy are neglected.

\begin{table}[]
  \centering
  \caption{Calculated reaction energies ($\Delta E = \sum\limits_{products}E-\sum\limits_{reactants}E$), independent equilibrium constants (300 K) and concentrations for Schottky disorder in \ce{CH3NH3PbI3}. 
For partial disorder the chemical potentials are taken to be pinned to the formation of \ce{PbI2} and \ce{CH3NH3I}, respectively.  
  The values of $K_C$ (\%) are normalised to the site fraction of vacancies, and $n$ refers to the associated vacancy defect concentration. }
    \begin{tabular}{lcccc}
    \hline
  Reaction  & $\Delta H_S$ (eV per defect)   & $K_C$  &  $n$ (cm$^{-3}$) \\
    \hline
    $\textit{nil} \rightarrow \ce{V_{MA}^/} + \ce{V_{Pb}^{//} + 3\ce{V_I}^{\bullet}} + \ce{MAPbI3}$  
    & 0.14 & 0.41 & $2\times10^{19}$  \\
    $\textit{nil} \rightarrow \ce{V_{MA}^/} + \ce{V_I}^{\bullet} + \ce{MAI}$  
    & 0.08 & 3.82 &  $2\times10^{20}$ \\
    $\textit{nil} \rightarrow \ce{V_{Pb}^{//} + 2\ce{V_I}^{\bullet}} + \ce{PbI2}$  
    & 0.22 & 0.02 &  $8\times10^{17}$ \\
  \hline
  \hline
    \end{tabular}%
  \label{table1}%
\end{table}

\begin{figure}[]
\centering
\vspace{0.5cm}
\includegraphics*[width=16.0cm]{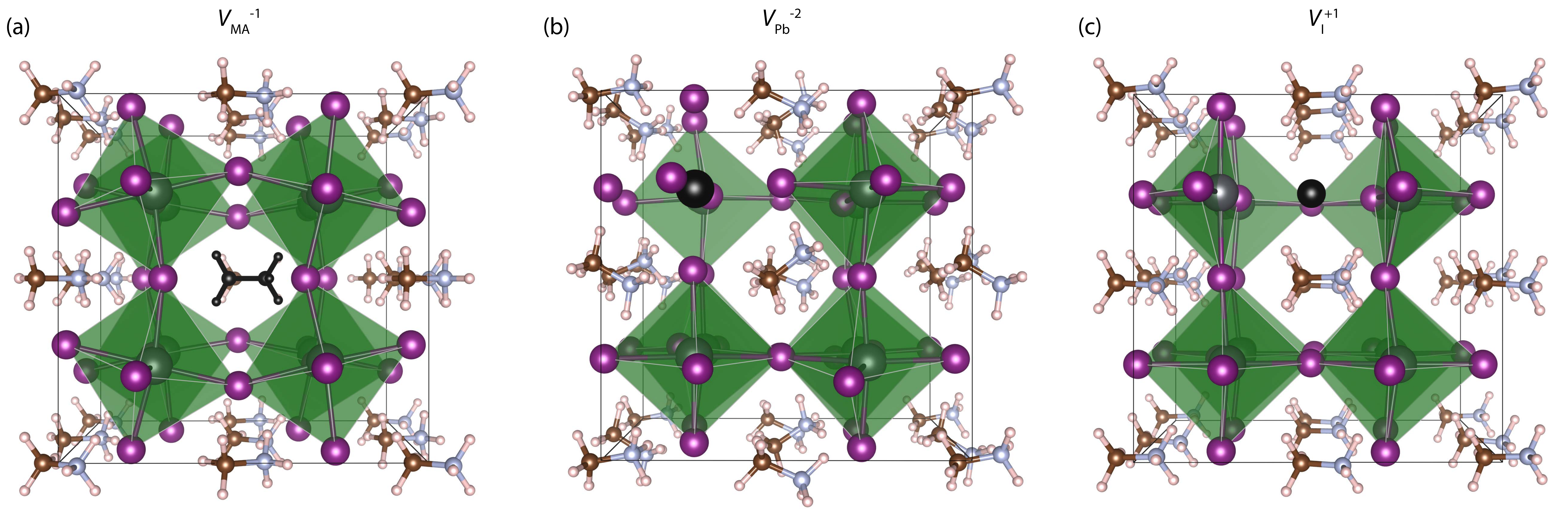}
\caption{Calculated local structure around the charged \ce{CH3NH3+}, \ce{Pb^2+} and \ce{I-} point defect vacancies in \ce{CH3NH3PbI3} that contribute to Schottky ionic disorder. 
The nominal vacancy site (missing chemical species) in shown in black for each case.
The dipole response to defect formation is driven by a complex combination of molecular reorientation and octahedral distortions (see Ref. \cite{frost2014molecular} for dynamic structural analysis).
Note that under conditions of charge and mass conservation  $[\ce{V_{MA}^/}]+[\ce{V_{Pb}^{//}]=3[\ce{V_I}^{\bullet}}]$.
}
\label{f1}
\end{figure}

The computed reaction energies -- a combination of the formation energy of the each of the individual charged point defects --  are summarised in Table 1. 
The local structures are drawn in Figure \ref{f1}.
The Schottky formation energy of 0.14 eV per defect is remarkably low, and corresponds to an equilibrium vacancy concentration of 0.4 \% at room temperature. 
In comparison, the reported Schottky formation energy for \ce{BaTiO3} is 2.29 eV per defect,\cite{smith-2000} which results in ppm equilibrium vacancy concentrations.
Interestingly, the reactions for partial Schottky formation are most favourable with respect to the loss of MAI, which at 0.08 eV per defect suggests that up to 4 \% of the \ce{CH3NH3} and \ce{I} sublattice will be vacant (in an open system). 
Such non-stoichiometric behaviour goes beyond the non-interacting point defect limit; hence, inter-defect correlations will be important to consider in future quantitative models.

The defect chemistry of this hybrid halide perovskite is unusual. 
It is common for wide band gap materials to favour ionic disorder (self-compensated arrangements of charged point defects) and low band gap materials to favour electronic disorder (a distribution of carriers in the valence and conduction bands). 
Here, ionic disorder is favoured despite the fact that electrons and holes are facile to form, with all of the defects investigated here being shallow donors (\ce{V_I}$^{\bullet}$) or acceptors (\ce{V_{Pb}}$^{//}$ and \ce{V_{MA}}$^/$).\cite{yin-063903} 
The self-regulation of equilibrium electron and hole concentrations will be provided by the formation
of charge-compensating lattice vacancies. 

One factor behind this behaviour is the lattice energy: in comparison to metal oxide perovskites, for halides the electrostatic potential of all lattice sites is reduced due to the lower formal oxidation states.\cite{frost2014atomistic} 
The electrostatic contribution to the vacancy formation energy varies with the square of the charge, e.g. $q^2$ = 4 for oxides (O$^{2-}$); 1 for halides (I$^{1-}$). 
Hence for the metal halides both anion and cation vacancies can form with high probability due to a decrease in the chemical bond strength.
While point defects, formed due to configurational entropy, facilitate significant equilibrium hole densities of up to 10$^{18}$cm$^{-3}$ in thin-film photovoltaic absorbers such as \ce{Cu2ZnSnS4},\cite{chen-1522} Schottky disorder in \ce{CH3NH3PbI3} limits their formation as it provides a route to minimise the free energy of the crystal \textit{without} generating charge carriers.

It is now well established that carrier diffusion lengths in hybrid perovskite thin films are long ($> 1 \mu m$).\cite{snaith-6156}
The effective carrier mobility has been estimated to be ca. 20 $cm^2V^{-1}s^{-1}$.\cite{leijtens2014electronic}
The contribution of Schottky disorder to electron transport (carrier lifetime), must therefore not be detrimental. 
This can be understood by the low charge and high dielectric constant of these materials, which limits the cross-section associated with ionised impurity scattering. 
Clustering of the vacancies into charge neutral combination will further suppress this process. In addition, due to the absence of mid-gap defect states, Shockley-Read-Hall electron-hole recombination is not expected for this type of disorder. 

The behaviour reported here has important implications for the application of hybrid perovskites in photovoltaics:
(i) The stoichiometric hybrid perovskites can simultaneously be highly defective and electronically benign, with the low carrier concentrations ensuring effective Fermi-level splitting for operation of a $p-i-n$ photovoltaic device.\cite{edri2014elucidating} The combination of high carrier mobility and built-in electric fields can efficiently drift photo-generated electrons and holes towards the $p$ and $n$ contacts; if the background carrier concentration was too high the $i$ region would not be fully depleted. 
(ii) Empirically a synthesis route rich in MAI precursors has been adopted,\cite{snaith-1511} which increases the chemical potentials of MA and I; hence, suppressing the partial Schottky disorder proposed in Reaction 2. 
The thermodynamic balance for disproportionation into the binary iodides is delicate, but should be preventable with appropriate encapsulation, which avoids loss of the more volatile \ce{MAI} component.
(iii) The high concentration of vacancies on all sites will facilitate mass transport, supporting the ionic conductivity evidenced in impedance spectroscopy,\cite{laurie} and supporting it as one of the possible causes of hysteresis\cite{snaith-1511} in the current-voltage behaviour and the giant dielectric constant at low frequencies.\cite{perez-2390} 
The perovskite crystal structure can support vacancy-mediated diffusion on each of the lattice sites.\cite{islam2000ionic} 
%

In summary, the unusual defect chemistry of \ce{CH3NH3PbI3} identified here is key to its success as an intrinsic photovoltaic material.
The first report by Weber on this material in 1978,\cite{weber1978ch3nh3pbx3} concluded `the compounds show intense colour, but there is no significant conductivity', which our model can now explain.
 If the self-regulation predicted for the stoichiometric material  could be overcome, either through extrinsic doping or kinetic control of non-stoichiometry, the extension of hybrid perovskites to a wider range of photovoltaic architectures would be possible.    


\textbf{Theoretical Methods}
The total energy of bulk and defective \ce{CH3NH3PbI3} 
was calculated in a 96 atom pseudo-cubic perovskite ($2\times2\times2$) supercell with a $\Gamma$-centred $3\times3\times3$ Monkhorst-Pack special $k$-point grid (400 eV plane wave cut-off), using a set-up previously reported.\cite{brivio-042111, frost2014atomistic}
Lattice-dynamics calculations were performed to ensure the absence of imaginary zone-centre phonon frequencies in the structures and supercells considered.
The main approximation is the supercell size, in particular relating to the long-range order of the methylammonium ions. 

The formation energy of the individual charged defects is defined as $E_{supercell}^{defective}-E_{supercell}^{bulk}$. The total energy of the charged defective systems
were corrected\cite{freysoldt-253} to account for: 
alignment of the electrostatic potential between the bulk and the defective supercells;
finite size effects in the calculation of charged impurities; 
band filling by defect levels resonant in the band edges. 
The static dielectric constant of 24.1 was employed in the calculations,\cite{brivio-2014} which 
include the electronic and vibrational response of the system, but 
excludes the rotational response of the dipolar molecules, which can occur at lower frequencies. 

All structures and energies were calculated using Kohn-Sham density functional theory (in the code \textit{VASP}). 
Interactions between the core and valence electrons is described within the PAW method\cite{vasp1}
including scalar-relativistic corrections. 
Spin-orbit coupling was not included, which is not expected to affect the defect structures -- the unoccupied Pb 6p conduction bands are most perturbed --  but it will be essential for quantitative defect spectroscopy.

The PBEsol exchange-correlation functional was employed.\cite{pbesol} 
PBEsol is a revision of the PBE functional, specifically tailored for solids, and has been shown to yield structural data in agreement with experiment.\cite{brivio-042111} 
This functional reproduces the structure of common `London dispersion' correction functionals without the addition of an empirical potential. 
Due to the ionic nature of the hybrid perovskite system, secondary polarisation 
is a minor effect.

\acknowledgments
We acknowledge membership of the UK's HPC Materials Chemistry Consortium, which is funded by EPSRC (EP/L000202/1)
and the Materials Design Network. 
A.W. acknowledges support from the ERC (Grant 277757) and EPSRC (EP/K016288/1 and  EP/M009580/1). 
D.O.S. acknowledges support for the NSF of China (11450110056). 
S.C. and X.G. are supported by Special Funds for Major State Basic Research, and NSFC (61106087, 91233121). 
The work at NREL is funded by the U.S. Department of Energy under Contract No. DE-AC36-08GO28308. 

\bibliography{library_aw}

\end{document}